\documentclass[useAMS,onecolumn]{mn2e}
\usepackage{psfig}
\newif\ifAMStwofonts

\usepackage{bm}

\title[The speed of gravity versus the speed of light]
{Note on the relationship between the speed of gravity and light in the bi-metric theory of gravity}
\author[Sergei M. Kopeikin]
       {Sergei M. Kopeikin\\ \\
       Department of Physics \& Astronomy, University of Missouri-Columbia, Columbia, MO 65211, USA}
\date{Accepted.............2006... ;  
      Received ............2006... ; 
      in original form ...........2006... }
\pubyear{2006}

\def\LaTeX{L\kern-.36em\raise.3ex\hbox{a}\kern-.15em
    T\kern-.1667em\lower.7ex\hbox{E}\kern-.125emX}

\begin{document}
\Large
\label{firstpage}

\maketitle
\begin{abstract}
Relationship between the speed of gravity $c_{\rm g}$ and the speed of light $c_{\rm e}$ in the bi-metric theory of gravity is discussed. We reveal that the speed of light is a function of the speed of gravity which is a primary fundamental constant. Thus, experimental measurement of relativistic bending of light propagating in time-dependent gravitational field directly compares the speed of gravity versus the speed of light and tests if there is any aether associated with the gravitational field considered as a transparent `medium' with the constant refraction index.
\end{abstract}
\begin{keywords}
gravitation -- gravitational waves -- relativity -- instrumentation: interferometers 
\end{keywords}

Any assumption that the fundamental speed (of light) in Maxwell's theory is different from the fundamental speed (of gravity) in Einstein's theory of general relativity inevitably leads to a bi-metric  theory of gravity operating with two metric tensors. One of them, $g_{\alpha\beta}$, introduces a gravity null cone along which weak gravitational waves propagate with speed $c_{\rm g}$, while the other, $\tilde g_{\alpha\beta}$ makes a light null cone along which electromagnetic waves propagate with speed $c_{\rm e}$. Had the difference between the two null cones existed the exact Lorentz symmetry of the Einstein gravity field equations would be broken with respect to the Lorentz transformation present in the Maxwell theory. The Lorentz groups of the Einstein and Maxwell equations are parametrized respectively by the speed of gravity $c_{\rm g}$ and the speed of light $c_{\rm e}$. According to Einstein's general principle of relativity (Landau \& Lifshitz 1971) these two speeds must be identical and equal to the fundamental speed $c$ of the Minkowski space-time. Thus, experimental measurement of the difference between the speed of light and that of gravity is crucially important for testing the Lorentz invariant property of gravity, that is for experimental confirmation that gravity do obeys the general principle of relativity. Various scenarios of the spontaneously broken violations of the Lorentz invariance of the gravitational field at high energies are given by Jacobson, Liberati \& Mattingly D. (2006). We discuss such violations at low energies in case of weak gravitational field in the solar system (Kopeikin 2004, 2006; Kopeikin \& Fomalont 2006). 

To model presumable difference between the two speeds, $c_{\rm e}$ and $c_{\rm g}$, it is convenient to employ the bi-metric theory of gravity proposed by Carlip (2004) which is general enough to make our analysis as complete as possible. This theory introduces the two metric tensors and defines relationship between the speed of light $c_{\rm e}$ and the speed of gravity $c_{\rm g}$ as
\begin{equation}
\label{1}
c_{\rm e}=c_{\rm g}\sqrt{1-\epsilon}\;,\qquad\qquad(\epsilon\le 1)
\end{equation}
where $0\leq\epsilon\leq 1$ is a sliding parameter characterizing the degree of violation of the Lorentz invariance of gravity with respect to light that is the difference between the gravity and light null cones. Quantity $n\equiv 1/\sqrt{1-\epsilon}$ can be viewed as a constant refraction index of vacuum filled with the gravitational field. In order to measure $\epsilon$ one has to conduct gravitational experiment in which the post-Newtonian terms of both metrics, $g_{\alpha\beta}$ and $\tilde g_{\alpha\beta}$, interfere \footnote{The post-Newtonian terms define deviation of the metric tensor from the Minkowski space-time caused by the presence of the gravitational field.}. We have discovered (Kopeikin 2001, Fomalont \& Kopeikin 2003, Kopeikin \& Ni 2006)  that practical measurement of the parameter $\epsilon$ can be rendered in gravitational light-ray deflection experiments in which light propagates through time-dependent gravitational field of a moving massive body. However, proper physical interpretation of this measurement requires to make choice of a specific system of units associated with the existence of the two metric tensors (Kopeikin 2005). To this end one can choose either gravitodynamic, $c_{\rm g}=1$, or electrodynamic system, $c_{\rm e}=1$, of units depending on a particular theoretical framework used in the data processing algorithm.

If one assumes the gravitodynamic system of units with $c_{\rm g}=1$ (Carlip 2004), then, the speed of light $c_{\rm e}$ will be measured, and it is expressed in terms of the parameter $\epsilon$ as $c_{\rm e}=\sqrt{1-\epsilon}$. On the other hand, if one assumes the electrodynamic (metric) system of units (Kopeikin 2001, 2004, 2006; Fomalont \& Kopeikin 2003; Kopeikin \& Fomalont 2006) with the speed of light $c_{\rm e}=1$, then the speed of gravity is measured, and it is expressed in terms of the parameter $\epsilon$ as  $c_{\rm g}=1/\sqrt{1-\epsilon}$. We emphasize that neither the speed of light nor the speed of gravity can be measured in gravitational experiments independently of the assumption of the system of units used for measuring distances in the solar system. At any rate, irrespective of the choice of the system of units the primary measured parameter of the bi-metric theory of gravity is 
\begin{equation}
\label{2}
\epsilon=1-\left(\frac{c_{\rm e}}{c_{\rm g}}\right)^2\;.
\end{equation}
Notice that this important parameter of the bi-metric theory of gravity is missed in the Nordtvedt-Will PPN formalism (Will 1993, 2005) because it has unsatisfactory theoretical treatment of the problem of propagation of light in alternative theories of gravity (Kopeikin 2006, Kopeikin \& Fomalont 2006, Kopeikin \& Ni 2006) along with some other subtle theoretical flaws like violation of the gauge-invariance and inadequate treatment of scalar and vector fields entering the PPN metric tensor (Kopeikin \& Vlasov 2004) \footnote{See the textbook by Ciufolini \& Wheeler (1995) for other criticism of the PPN formalism}. Hence, the PPN postulates $\epsilon=1$ and eliminates it from the set of the legitimate PPN parameters.

We emphasize that the speed of gravity $c_{\rm g}$ is a primary fundamental constant in Carlip's bi-metric theory of gravity (Carlip 2004) which value does not depend on the choice of a frame of reference. This property of the speed of gravity in Carlip's theory makes it clear that any reference frame can be used to measure its value with respect to the speed of light. Therefore, we do not specify a particular frame in which the equation (\ref{4}) is valid since it is gauge-invariant.  The speed of gravity $c_{\rm g}$ enters the gravitational metric $g_{\alpha\beta}$ as well as its first and second time derivatives, that is in the Christoffel symbols and the Riemann tensor respectively. Will (2003) postulates that the speed of gravity enters only to the second time derivatives of the metric tensor that is to the Riemann tensor while the Christoffel symbols depend on the speed of light. This makes the Christoffel symbols electromagnetic-field dependent which is erroneous postulate (Kopeikin 2006; Kopeikin \ Fomalont 2006) contradicting Einstein's geometric approach to gravity field theory (Landau \& Lifshitz 1971). Had $c_{\rm g}=\infty$ the time derivatives of the metric tensor were totally suppressed and their impact on the relativistic deflection and/or time delay of light could not be observed in the gravitational experiments conducted in time-dependent gravitational fields  (Kopeikin 2004, 2006; Kopeikin \& Fomalont 2006). The speed of light $c_{\rm e}$ in Carlip's bi-metric theory (Carlip 2004) enters through the coordinate of the light particle (photon) moving in the gravitational field. In contrast to the speed of gravity, the speed of light $c_{\rm e}$ is frame-dependent and refers to the constant speed of gravity $c_{\rm g}$ via parameter $\epsilon$ in equation (\ref{1}).

Gravitational experiment to measure the speed of gravity with respect to the speed of light has been proposed in our paper (Kopeikin 2001) and completed in September 2002 by Fomalont and Kopeikin (2003). The basic idea of the experiment was to measure the retarded position of Jupiter on its orbit by observing the tangential component of the gravitational deflection of light of a quasar caused by the moving gravitational field of Jupiter. In case of $c_{\rm g}=c_{\rm e}$ the tangential component is absent and the deflection of light by a moving object (Jupiter) taken on its orbit at the retarded instant of time, is purely radial as seen in the plane of the sky (Kopeikin 2001, 2004, 2006; Fomalont \& Kopeikin 2003; Kopeikin \& Fomalont 2006). Non-radial component of the gravitational deflection of light can be also treated either as gravitomagnetic dragging of light ray caused by translational current of matter associated with the orbital motion of the light-ray deflecting body (Kopeikin 2004, 2006; Kopeikin \& Fomalont 2006; Sereno 2005a,b) or as the aberration of gravity versus the aberration of light (Fritelli 2003, Kopeikin 2006, Kopeikin \& Fomalont 2006). 

In order to interpret the results of the Jovian experiment properly one must understand how the theoretical model tells us the coordinates of Jupiter are measured (Kopeikin 2005). 
Coordinates ${\bf x}_J$ and velocity ${\bf v}_J$ of Jupiter can be `theoretically observed' in the bi-metric theory of gravity in two ways - with ranging process based on propagation of either electromagnetic or gravitational wave signals \footnote{Presently, only light/radio ranging can be rendered practically.}. The corresponding equations describing the ranging process in the first relativistic approximation result from the Maxwell and gravity field equations of the bi-metric theory (Carlip 2004). They read respectively as follows (Kopeikin \& Ni 2006)
\begin{eqnarray}
\label{3}
t-t_e&=&\frac{1}{c_{\rm e}}|{\bm x}-{\bm x}_J(t_e)|\;,\qquad\mbox{(light null-cone ranging)}\\
\label{4}
t-t_g&=&\frac{1}{c_{\rm g}}|{\bm x}-{\bm x}_J(t_g)|\;,\qquad\mbox{(gravity null-cone ranging)}
\end{eqnarray}
where $t$ and ${\bm x}$ are time and space coordinates of emitter, $t_e$ and $t_g$ are respectively the instants of time when the `emitted' electromagnetic and gravitational signal reach Jupiter. We emphasize that in the ranging measurement only time intervals and the speed of the signal used for the measurement are known, so that coordinates of the ranging body (Jupiter) are expressed in terms of them. 

If one postulates (Carlip 2004) that the speed of gravity $c_{\rm g}=1$ \footnote{Any other constant numerical value of $c_{\rm g}$ works in the same way and leads to the same physical conclusions. Carlip (2005) erroneously believes that theoretical interpretation of the Jovian experiment (Fomalont \& Kopeikin 2003) depends on whether $c_{\rm g}$ appears explicitly in theoretical equations or not.} then the speed of light $c_{\rm e}$ is a measurable quantity and, thus, can not be used in the definition of ranging distances in the solar system. This crucial point is clearly understood, for example, by Wolf \& Petit (1997) in contrast to Carlip (2004, 2005) and Will (2005) where the speed of light $c_{\rm e}$ is inconsistently
 used both as a measured parameter and as a constant for measuring ranging distances. 
In the system of units with $c_{\rm g}=1$ the ranging coordinates of Jupiter ${\bm x}_J$ are mathematically expressed in terms of the product of the theoretically known speed of gravity $c_{\rm g}=1$ and the time interval $t-t_g$ from equation (\ref{4}), that is they implicitly depend on the speed of gravity $c_{\rm g}=1$ as shown in our paper (Kopeikin 2005). Hence, measuring the speed of light $c_{\rm e}$ in the system of units with $c_{\rm g}=1$  tests whether the speed of gravity $c_{\rm g}=1$ or not. Currently gravitational waves are not directly measurable with gravitational wave detectors. Hence, we can not use these detectors for direct practical measurement of distances to astronomical bodies in the solar system. None the less, we can use gravitational bending of light by a moving mass in order to measure its retarded coordinate defined in the bending on the basis of the gravity null-cone ranging equation (\ref{4}), and to compare it with the coordinate of the mass obtained on the basis of the light/radio ranging equation (\ref{3}) (Kopeikin 2006, Kopeikin \& Fomalont 2006). 

Indeed, the speed of light $c_{\rm e}$ is precisely known and fixed in SI system of units by the value of 299792458 m/s. Specifically this value of the speed of light was used in the Jovian experiment (Fomalont \& Kopeikin 2003) as a reference speed for measuring the speed of gravity $c_{\rm g}$. It is important to realize that both the `speed of light' (Carlip 2004, Will 2005) and the `speed of gravity' (Kopeikin 2001, Fomalont \& Kopeikin 2003) interpretations of the Jovian experiment are mathematically equivalent but making particular emphasis on the `speed of light' interpretation (Will 2005) conceals the true physical meaning of the Jovian experiment as test of the Lorentz-invariance of gravity with respect to that of light. Our finding (Fomalont \& Kopeikin 2003) that $c_{\rm g}=c_{\rm e}$ means that the gravitational field obeys the Einstein general principle of relativity and does propagate with finite speed (Kopeikin 2006, Kopeikin \& Fomalont 2006). 

Fomalont \& Kopeikin (2003) experiment measured the time delay $t-t_g$ in the left side of equation (\ref{4}) by observing the time-dependent post-Newtonian correction to the Shapiro time delay associated with the gravity null-cone retarded position of Jupiter. This time delay was compared with the time delay $t-t_e$ in equation (\ref{3}) that is based on the radio ranging and VLBI measurements of satellites orbiting Jupiter and inferred from JPL ephemeris (Standish, private communication). This comparison allowed us to evaluate the speed of gravity $c_{\rm g}$ with respect to the speed of light $c_{\rm e}$ (Fomalont \& Kopeikin 2003). We have conducted similar experiment in the field of the sun as it passed in front of quasar 3C279 (Kopeikin 2006, Kopeikin \& Fomalont 2006). Results of the new experiment will be published somewhere else. 

Scrutiny experimental study of tiny relativistic effects in propagation of light through time-dependent gravitational fields can be achieved in space-borne laser experiments (Ni 2005; Kopeikin \& Ni 2006), from on-board of space astrometric missions Gaia (Perryman 2005) and SIM (Shao 2004), and at super-advanced ground-based radio observatories like the Square Kilometer Array (Schilizzi 2004). Success of these and other projects can significantly advance our knowledge of fundamental laws of gravity and help us to connect gravitational physics at the quantum scale with the geometric tenets of general theory of relativity. 

This work has been supported by grants of the Eppley Foundation for Research (New York) and the Research Council of the University of Missouri-Columbia. We thank Steve Carlip for discussion.

\end{document}